\documentclass{webofc}
\usepackage[varg]{txfonts}   
\usepackage{braket}
\usepackage{multirow}
\usepackage{dsfont}
\usepackage{mathtools}
\usepackage{xcolor}
\usepackage[normalem]{ulem}

\begin{document}
\title{Reply to E. Shuryak's Comments on "Three regimes of QCD"}
%
%

\author{\firstname{L. Ya. } \lastname{Glozman}\inst{1}\fnsep\thanks{\email{leonid.glozman@uni-graz.at}} 
}

\institute{Institute of Physics, University of Graz, A-8010 Graz, Austria
          }

\abstract{ In his recent Comments E. Shuryak reiterates old,
unfortunately misleading arguments in favor of deconfined
Quark-Gluon Plasma (QGP) immediately above the chiral restoration
pseudocritical temperature. In a Comment devoted to our view
of QCD at high temperatures he does not address and even 
mention the essence of our arguments. In recent years a new
"hidden" symmetry in QCD was discovered. It is a symmetry of the
electric sector of QCD, that is higher than the chiral symmetry
of the QCD Lagrangian as the whole. This symmetry was clearly observed
above $T_c$ in spatial correlators and very recently also in time correlators.
The latter correlators are directly related to observable spectral
density. Then in a model-independent way we conclude that degrees
of freedom in QCD above $T_c$, but below roughly $3T_c$, are chirally
symmetric quarks bound by the chromoelectric field into color-singlet
compounds without the chromomagnetic effects. This regime of QCD has been referred to as a Stringy Fluid since such objects are very reminiscent of
strings.At higher temperatures there is a very smooth transition to the
partonic degrees of freedom, i.e. to the QGP regime. Here we will
address some of the points made by Shuryak.}
\maketitle
\section{Introduction}
\label{intro}

In  Comment \cite{S} on our recent overview talk \cite{talk}
Edward Shuryak tries to defend the "orthodox view" on QCD
phase diagram at high temperatures against our "unorthodox"
position. It is more or less understood that QCD at low temperatures is
a Hadron Gas, while at a very high temperatures a partonic description
should be adequate (the QGP). It is a  long way
from one regime to another. The "orthodox view" defended by Shuryak
assumes that there is an abrupt transition from the hadron degrees of
freedom below $T_c$ to the partonic degrees of freedom above $T_c$.

Shuryak does not address the key point of our view
that is based on a new symmetry of the electric sector of QCD and its
observation in lattice calculations in spatial and time correlators 
above $T_c$.
Unfortunately he does not cite the relevant papers
that should be mentioned since a number of people were involved
into this "orthodox $\rightarrow$ unorthodox" evolution. 

An incentive to discovery of a new symmetry was the observation
of a large degeneracy of hadrons upon artificial amputation of
the lowest modes of the Dirac operator in lattice simulations,
a degeneracy that is much larger than it would be implied by
the chiral symmetry of the QCD Lagrangian \cite{D1,D2,D3,D4}. 
The symmetry group responsible for this degeneracy, called the
chiral-spin symmetry $SU(2)_{CS}$ and its flavor extension $SU(2N_F)$,
as well as the observation that it is a symmetry of the electric sector
in QCD, was obtained in refs. \cite{G1,G2}. Then it was predicted
that such a symmetry should exist above $T_c$ \cite{G3}.
Detection of this symmetry in spatial correlators in the $T_c - 3T_c$ 
interval with the chirally-symmetric Domain Wall Dirac operator
within the $N_F=2$ QCD using the JLQCD ensembles was done in Refs.
\cite{R1,R2}. The same symmetry was very recently  seen
immediately above $T_c$ also in t-correlators, that are directly
connected to measurable spectral density \cite{R3}. Finally,
since this symmetry is also a symmetry of the chemical potential
in the QCD action, it should also persist at finite chemical
potentials \cite{G4}.

 Here we will not address all statements made in \cite{S} and
 only shortly comment some of them since they have been used for
 a long time to justify the existence of QGP above $T_c$.
 
 \section{Section II}
 
 In section II Shuryak discusses the screening of the
 color charge at high temperatures. He pioneered the perturbative
 calculations at high T and obtained the Debye screening
 of the electric field like in usual electric plasma \cite{SD}.
 Hence, he concludes  - deconfinement, that is also observed on the lattice 
 via the flattering of
 the potential between the static color sources.
 
 In our lattice calculations we clearly see the chiral-spin and $SU(4)$
 symmetries in  spatial and time correlators at temperatures
 $T_c - 3T_c$ \cite{R1,R2,R3}. Given this we conclude in a model-independent way
 that elementary degrees of freedom are the color-singlet compounds
 that consist of chirally symmetric quarks that are bound by the
 chromoelectric field without the chromomagnetic effects. Hence the name
 - a Stringy Fluid. This result implies that there is no
 Debye screening of the electric field in this temperature range. Just opposite,
 it is a magnetic field that is screened. Consequently physics in this
 highly nonperturbative regime is radically different as compared
 to perturbative predictions that are perhaps valid at huge temperatures.
 
 With regard to  the Polyakov loop and  the potential
 between the static sources we will simply cite two paragraphs from
 the introduction in Ref. \cite{R2}:

\bigskip
"The expected confinement-deconfinement transition turned out to be more intricate
to define. Such a transition was historically assumed to be associated with a different
expectation value of the Polyakov loop \cite{P,L} below and above the critical
temperature $T_c$. In pure $SU(3)$ gauge theory the Polyakov loop is connected
with the $Z_3$ center symmetry and indeed a sharp first-order phase transition
is observed \cite{Kaczmarek:2002mc}, which indicates that the relevant degrees of freedom below and above
$T_c$ are different. Still, one may ask whether this $Z_3$ transition is really connected with deconfinement in 
a pure glue theory. Traditionally the answer was affirmative, because the expectation value of the 
Polyakov loop can be related to the free energy of a static quark source. If this
energy is infinite, which corresponds to a vanishing Polyakov loop,
then we are in a confining mode, while deconfinement should be
associated with a finite free energy, i.e., a non-zero Polyakov loop.
However, this argumentation is  self-contradictory because a criterion for deconfinement in pure
gauge theory, i.e., deconfinement of gluons, is reduced to deconfinement
of a static charge (heavy quark), that is not part of the pure glue theory. 
The Polyakov loop is a valid order parameter but strictly speaking its relation to confinement is an assumption. And
indeed, just above the first-order $Z_3$  phase transition the energy and
pressure are quite different from the Stefan-Boltzmann limit which is associated
with free deconfined gluons \cite{b}.

In a theory with dynamical quarks the first-order phase transition is
washed out and on the lattice one observes a very smooth increase
of the Polyakov loop \cite{Petreczky:2015yta}. The reason for that behavior is rather clear:
in a theory with dynamical quarks there is no $Z_3$ symmetry and the
Polyakov loop ceases to be an order parameter.
Considering the finite energy of a pair of static quark sources (Polyakov loop correlator)
the resulting string breaking potential is due to vacuum loops of light quarks that combine 
with the static sources to a pair of heavy-light mesons."

\bigskip
Concerning the equation of state, that is also discussed by Shuryak in
this section, it is very well established that a simple hadron
resonance gas model does describe the growth of pressure and energy density
up to temperatures slightly above $T_c$. Here both the energy and the pressure
are very far from the Stefan-Boltzmann limit, that is associated with
Eqs. (2)-(3) of Shuryak's Comments. Then, a rapid growth begins that
qualitatively coincides with the $T_c - 3T_c$ interval of Stringy Fluid.
But still, in this interval both pressure and energy density are quite
 far from the Stefan-Boltzmann limit \cite{Bazavov:2017dsy}. Above
 these temperatures a slow evolution to the Stefan-Boltzmann limit
 is observed that could perhaps be described within the hard thermal loops
 approach \cite{PB,BR}.
 
 It is an interesting and very important question to clarify physics
 of the smooth transition from condensed strings (the Stringy Fluid regime)
 to partonic degrees of freedom (the QGP regime).

\section{Section III}

Here Shuryak addresses our truncation studies \cite{D1,D2,D3,D4}
and insists that they are compatible with the instanton liquid model
developed by him and others in the past \cite{ILM}.

Here we observed that  after removal of order 10 lowest modes
of the Dirac operator from the quark propagators both 
$SU(2)_L \times SU(2)_R$ and $U(1)_A$ symmetries are restored in the
hadron spectrum. This result could  indeed be explained by the assumption
that both chiral symmetries breakings are due to instantons in the
liquid phase. However we observed a much larger symmetry, than simply
the chiral symmetry of QCD, after
truncation of the lowest modes. Then, given the symmetry classification
of the QCD Lagrangian, we conclude that while the confining chromoelectric
interaction is distributed over all modes of the Dirac operator, the
magnetic interaction is localized at least predominantly in the near zero modes \cite{G1,G2}. Such a peculiar result cannot be derived within the instanton
liquid model and requires dynamics that incorporates at the same
time both confinement and chiral symmetry breaking. One can speculate
that this physics could be obtained assuming that the global gauge configurations that should contain confinement physics, include
local topological fluctuations that contribute into the near-zero
modes of the Dirac operator.

\section{Section IV}

We will not comment here some ideas related to monopoles
and magnetic screening, addressed by Shuryak in  subsection A.
In subsections B-E he discusses our results on spatial correlators
as well as calculates within a model screening masses of spatial correlation
functions. He  claims that our statements are the same
as  those
made by DeTar and Kogut long ago \cite{DTK}, who first studied
screening masses of spatial correlators. Our conclusions are
based however not on screening masses but on observed  $SU(2)_{CS}$ and $SU(4)$ symmetries of the
spatial correlators at $T_c - 3T_c$ \cite{R1,R2} \footnote{Edward is very
well aware of this since it was discussed with him many times. He
cannot do anything with these symmetries. Then a very well known in politics
method is used: to ascribe to the opponent on public a statement that he
did not make and  attack this statement.}.
Observation of the
$SU(2)_{CS}$ and $SU(4)$ symmetries in spatial correlators does imply
in a model-independent way that it is a chromoelectric flux between quarks
which binds chirally symmetric quarks at these temperatures, and not a "magnetic confinement"
related to spatial Wilson loops of a pure glue theory (which is a model). We also note that
the highly nontrivial symmetries seen in spatial correlators in
the temperature range $T_c - 3 T_c$ cannot be obtained
via the dimensional reduction. In addition, these symmetries smoothly
disappear at higher temperatures, leaving intact only chiral symmetries.
 $SU(2)_{CS}$ and $SU(4)$ symmetries cannot be related to 3d physics,
since otherwise they would persist at higher temperatures, which is not observed. 

Concerning the screening masses, often fitted in the lattice papers
and then compared with the perturbation theory predictions,
we intentionally avoided this in our study. The reason is that the
screening mass assumes existence of a pole and consequently the $\exp(-\mu z)$
asymptotics. Within the perturbation theory there are no bound states
and the asymptotics is $\exp(-m z)/z$ and $\exp(-m z)/z^2$ depending
on  quantum numbers \cite{R2}. It is a  misleading procedure
to compare $\mu$ extracted from the exponential fit with the value $m$ obtained
within perturbation theory.

\end{document}